\documentclass[sigconf, nonacm]{acmart}

\usepackage{multirow}
\usepackage{balance}

\copyrightyear{2025}
\acmYear{2025}
\setcopyright{rightsretained}
\acmConference[RecSys '25]{Proceedings of the Nineteenth ACM Conference on Recommender Systems}{September 22--26, 2025}{Prague, Czech Republic}
\acmBooktitle{Proceedings of the Nineteenth ACM Conference on Recommender Systems (RecSys '25), September 22--26, 2025, Prague, Czech Republic}\acmDOI{10.1145/3705328.3748130}
\acmDOI{10.1145/3705328.3748130}
\acmISBN{979-8-4007-1364-4/2025/09}

\begin{document}

\title[Industry Insights from Comparing Deep Learning and GBDT Models for E-Commerce LTR]{Industry Insights from Comparing Deep Learning and GBDT Models for E-Commerce Learning-to-Rank}

\author{Yunus Lutz}
\orcid{0009-0001-1477-4547}
\email{yunus.lutz@otto.de}
\affiliation{%
  \institution{OTTO (GmbH \& Co. KGaA)}
  \city{Hamburg}
  \country{Germany}
}

\author{Timo Wilm}
\orcid{0009-0000-3380-7992}
\email{timo.wilm@otto.de}
\affiliation{%
  \institution{OTTO (GmbH \& Co. KGaA)}
  \city{Hamburg}
  \country{Germany}
}

\author{Philipp Duwe}
\orcid{0009-0001-1165-1832}
\email{philipp.duwe@otto.de}
\affiliation{%
  \institution{OTTO (GmbH \& Co. KGaA)}
  \city{Hamburg}
  \country{Germany}
  \vspace{0.6cm}
}

\acmArticleType{Research}

\ccsdesc[500]{Applied computing~Online shopping}
\ccsdesc[500]{Information systems~Recommender systems}
\ccsdesc[500]{Information systems~Learning to rank}

\keywords{learning to rank, search systems, recommender systems, e-commerce, gbdt, deep learning, online evaluation}

\begin{abstract}

In e-commerce recommender and search systems, tree-based models, such as LambdaMART, have set a strong baseline for Learning-to-Rank (LTR) tasks. Despite their effectiveness and widespread adoption in industry, the debate continues whether deep neural networks (DNNs) can outperform traditional tree-based models in this domain. To contribute to this discussion, we systematically benchmark DNNs against our production-grade LambdaMART model. We evaluate multiple DNN architectures and loss functions on a proprietary dataset from OTTO and validate our findings through an 8-week online A/B test. The results show that a simple DNN architecture outperforms a strong tree-based baseline in terms of total clicks and revenue, while achieving parity in total units sold.
\end{abstract}

\maketitle
\makeatletter{\renewcommand*{\@makefnmark}{}
\footnotetext{© Yunus Lutz, Timo Wilm, and Philipp Duwe 2025.
This is the author’s version of "Industry Insights from Comparing Deep Learning and GBDT Models for E-Commerce Learning-to-Rank". It is posted here for your personal use. Not
for redistribution. The definitive version of record was accepted for publication in the
19th ACM Conference on Recommender Systems (RecSys 2025). The final published version will be available at the ACM Digital Library: \\ ACM ISBN: 979-8-4007-1364-4/2025/09  \\ https://doi.org/10.1145/3705328.3748130\makeatother}

\section{Introduction}
LTR models are essential components of e-commerce recommender and search systems, where ranking quality directly influences user engagement and key business metrics. Gradient-boosted decision trees (GBDTs), particularly LambdaMART, have been the backbone of successful LTR systems due to their effectiveness and the availability of open-source implementations \cite{hu_unbiased_2019, magnani_semantic_2022, 10.1145/3077136.3080838}. 

An academic study reports the superiority of transformer-based DNN architectures over GBDTs for LTR tasks \cite{50030}.
Industry researchers confirm these findings, but could not provide online A/B test results due to scalability issues \cite{Buyl2023, DBLP:journals/corr/abs-2005-10084}.
Other practitioners report that a single-layer feed-forward neural network with a hidden size of 32 matches the performance of their GBDT baseline in an online A/B test, which raises questions about the strength of their baseline \cite{10.1145/3292500.3330658}.
These inconclusive results leave e-commerce practitioners uncertain about whether to adopt DNNs for their ranking tasks.

This work seeks to address this uncertainty by conducting a comprehensive empirical comparison of DNNs with our production-grade LambdaMART model in a large-scale e-commerce setting. We evaluate the impact of different model architectures and loss functions using a proprietary OTTO dataset and validate our findings through an 8-week online A/B test, providing industry practitioners with actionable insights for model selection in real-world e-commerce systems.

\section{Related Work}
The lack of open-source, large-scale e-commerce datasets for LTR has limited the evaluation of different machine learning models in this domain. Human-labelled datasets such as \textit{Web30K} \cite{DBLP:journals/corr/QinL13} and \textit{Yahoo! Learning to Rank Challenge} \cite{pmlr-v14-chapelle11a} are still the go-to datasets for LTR research \cite{10.1145/3340531.3412031, 10.1145/3477495.3531948}. These datasets are not representative of LTR datasets found in the industry, which are generally collected from interaction logs and contain implicit labels in the form of clicks or other signals. This makes practitioners question whether the research finding that DNNs can outperform GBDTs in LTR tasks, derived from these datasets, can be generalized to real-world e-commerce applications \cite{50030}.

The \textit{Baidu Unbiased LTR} \cite{zou2022large} dataset is gaining traction in research \cite{10.1145/3626772.3657892, li2023betterwebsearchperformance} because it is more realistic in size and structure. However, it still lacks important product features, which are crucial in e-commerce LTR applications \cite{10.1145/3077136.3080838}.

Therefore, e-commerce industry research comparing GBDTs and DNNs addresses this issue by evaluating models on proprietary interaction log data \cite{10.1145/3077136.3080838, Buyl2023, DBLP:journals/corr/abs-2005-10084}. Unfortunately, this research lacks comprehensive online evaluation through A/B tests, which is critical for assessing the real-world impact of LTR models, leaving practitioners with limited actionable insights.

Our work evaluates DNNs against a production-grade LambdaMART model, followed by an extensive online A/B test.

\section{Contributions}
This work provides a systematic evaluation of deep learning models for large-scale e-commerce LTR tasks and compares them to OTTO's production-grade LightGBM (LGBM) LambdaMART model. Our key contributions are the following:
\begin{itemize}
\item Extensive offline experiments on a large-scale proprietary dataset from OTTO demonstrate that DNNs can match or exceed the performance of tree-based models across multiple NDCG-based ranking metrics.
\item Three distinct DNN architectures and two loss functions are benchmarked, with a simple Two-Tower model achieving competitive performance compared to more complex models.
\item Offline results are validated through an 8-week online A/B test, showing that a simple DNN outperforms our production-grade LGBM model in engagement metrics, such as total clicks and revenue, while achieving parity in total units sold.
\end{itemize}

\section{Learning-to-Rank at OTTO}
\label{LTRAtOtto}

At OTTO, we address a contextualized LTR task that follows a candidate retrieval stage. Given a user request, the ranking system receives a list of \(n \in \mathbb{N}\) candidate products along with a contextual signal $\mathbf{c}$, which may include user behavior, query intent, or device information. The candidate set is denoted by \(\mathbf{p} = [\mathbf{p}_1, \mathbf{p}_2, \ldots, \mathbf{p}_n]\), where each \(\mathbf{p}_i\) represents a product retrieved by the upstream system. Each product $\mathbf{p}_i$ is described by a set of feature tensors, including \textbf{numerical} features $\mathbf{f}_i^\text{num}$, \textbf{categorical} features $\mathbf{f}_i^\text{cat}$, and \textbf{textual} features $\mathbf{f}_i^\text{text}$.
The LTR model assigns a relevance score $s_i$ to each product $\mathbf{p}_i$, conditioned on the context $\mathbf{c}$. Based on these scores, the candidate list $\mathbf{p}$ is re-ordered to produce the final ranked list. The objective is to present the user with the most relevant products at the top of the list.

For model training, we utilize real-world historical interaction data $(\mathbf{c}, \mathbf{p}, \mathbf{y}_c, \mathbf{y}_o)$, where users positively interacted with the product list $\mathbf{p} $ in context $\mathbf{c}$ either by clicking or ordering one or more items. Since multiple products can be interacted with, we represent clicks and orders as binary vectors $\mathbf{y}_c \in \{0,1\}^n$ and $\mathbf{y}_o \in \{0,1\}^n$, where each entry $\mathbf{y}_c^i, \mathbf{y}_o^i$ indicates whether product $\mathbf{p}_i$ was clicked or ordered in context $\mathbf{c}$.

\subsection{Feature Embeddings}
The features described in Section~\ref{LTRAtOtto} must be embedded into dense vector representations to serve as inputs for deep neural network architectures. \textbf{Numerical} features $f_{i,j}^\text{num}$ are inherently dense and include normalized attributes such as price, discount percentage, and historical engagement signals (e.g., clicks and orders). To address different distributional properties, power-law normalization is applied to right-skewed features \cite{10.1145/3292500.3330658}, while light-tailed features are standardized using z-score normalization.

For \textbf{categorical} features, we use embedding layers to map each categorical feature $f_{i,j}^\text{cat}$ to a dense vector representation. Specifically, for each categorical feature $j$ the embedding is computed as:
\begin{equation*}
  \mathbf{e}_j^\text{cat} = \text{Embedding}_j\left(f_{\cdot,j}^\text{cat}\right),
\end{equation*}
where $\mathbf{e}_j^\text{cat}$ is the resulting $d_j^\text{cat}$-dimensional embedding vector, and $j=1,\dots, C$ represents the $j$-th categorical embedding layer.

Furthermore, each \textbf{textual} feature $f_{i,j}^\text{text}$ for product $\mathbf{p}_i$ is represented as a bag-of-words vector $\mathbf{w}_j^{\mathbf{p}_i} = \left[w_1^{\mathbf{p}_i}, \dots, w_m^{\mathbf{p}_i}\right]_j$, where $j = 1, \dots, T$ denotes the $j$-th textual embedding layer, which can represent attributes such as product titles or descriptions. These bag-of-word vectors are fed into their corresponding embedding layer, where the embeddings of each word are summed:
\begin{equation*}
  \mathbf{e}_j^\text{text}=\sum\limits_{k=1}^{m} \text{Embedding}_j\left(w_k^{\mathbf{p}_i}\right) \in \mathbb{R}^{d^{\text{text}}_j}, \text{where } j=1,\dots, T.
\end{equation*} 
Finally, all embedding vectors are concatenated together to represent the final product embedding:
\begin{equation*}
  \mathbf{x}_{\mathbf{p}_i} = \text{concat}\Bigl(\Bigl[\left[f_{i,j}^\text{num}\right]_{j=1}^N, \left[\mathbf{e}_j^\text{cat}\right]_{j=1}^C, \left[\mathbf{e}_j^\text{text}\right]_{j=1}^T\Bigr]\Bigr) \in \mathbb{R}^{D},
\end{equation*}
where $D$ is the sum of all embedding vector dimensions and the total number of numerical features. The feature embedding $\mathbf{x}_c$ for context $\mathbf{c}$ is constructed similarly to $\mathbf{x}_{\mathbf{p}_i}$, but does not include numerical features.

\section{Architectures and Losses}
To explore the effectiveness of deep learning models in LTR tasks, we evaluate three distinct architectures.
All architectures use a backbone network with \(k = 1, \dots, N\) layers of the form:
\begin{equation*}
  B^{(k)}(\mathbf{z})
= \mathrm{LayerNorm}\Bigl(\mathrm{ReLU}\bigl(\mathrm{Dropout}\bigl(\mathbf{W}^{(k)} \mathbf{z} + \mathbf{b}^{(k)}\bigr)\bigr)\Bigr).
\end{equation*}
The layers are stacked in a recursive manner with skip connections:
\begin{equation*}
  \mathbf{z}_{k} 
= \mathbf{z}_{k-1} 
\;+\; B^{(k)}(\mathbf{z}_{k-1}),
\qquad
\in \mathbb{R}^{h}.
\end{equation*}
The input is projected linearly from $\mathbb{R}^{in}$ to  $\mathbb{R}^{h}$ before passing it to $B^{(1)}$, where $h$ is the hidden size. For brevity, we omit this notation. The final output of the backbone network is $f_{b}(\mathbf{z}) = \mathbf{z}_n \in \mathbb{R}^{h}$.

\subsection{Architectures}
We use the \textbf{Two-Tower} architecture \cite{45530}, where product features $\mathbf{x}_p$ are encoded by the backbone network, and context features $\mathbf{x}_c$ pass through a distinct linear layer: $\mathbf{h}_c = \mathbf{W}\mathbf{x}_c + \mathbf{b}\in \mathbb{R}^{h}$ and $\mathbf{h}_p = f_{b}(\mathbf{x}_p) \in \mathbb{R}^{h}$. The final scores $s_i$ for each product $\mathbf{p}_i$ are computed via a dot product between the two embeddings:
\begin{equation*}
 \mathbf{s} = [s_1,s_2,\dots,s_n] = \mathbf{h}_c^\top \mathbf{h}_p.
\end{equation*}
This architecture enables pre-computing the item embeddings $\mathbf{h}_p$, which allows for efficient scoring at inference time.

The \textbf{Cross-Encoder} model jointly encodes $\mathbf{x}_c$ and $\mathbf{x}_p$ with the backbone network and computes the final scores with a scoring layer $f_s$:
\begin{equation*}
  \mathbf{h}_z = f_b(concat([\mathbf{x}_p, \mathbf{x}_c])), \quad \mathbf{s} = f_s(\mathbf{h}_z) 
\end{equation*}

The \textbf{Transformer} model enhances the Cross-Encoder by using multi-head self-attention (MHSA) \cite{10.5555/3295222.3295349} without positional encodings to generate listwise contextual embeddings $\mathbf{h}_t$ for each product $\mathbf{p}_i$ in $\mathbf{p}$ \cite{46488}. The scores $\mathbf{s}$ are then calculated by combining $h_t$ and $h_z$ using a latent cross \cite{50030} and a scoring layer $f_g$:

\begin{equation*}
  \mathbf{h}_t = MHSA(\text{concat}([\mathbf{x}_p, \mathbf{x}_c])), \quad \mathbf{s} = f_g((1+\mathbf{h}_t)\odot \mathbf{h}_z)
\end{equation*}
The Cross-Encoder and Transformer models capture more complex feature interactions but introduce higher computational complexity.

\subsection{Losses}
The \textbf{RankNet (RN)} \cite{burges_learning_2005} and \textbf{Softmax Cross-Entropy (CE)} \cite{48321} losses are commonly used for LTR tasks. As discussed in Section~\ref{LTRAtOtto}, users can interact with multiple products. We modify the original RN loss ($\tilde{\mathcal{L}}_{\text{RN}}$) by dividing it by the number of positives  $P_n = \sum y^i$ in sample $n$:
\begin{equation*}
  \mathcal{L}_{\text{RN}}(\mathbf{s}, \mathbf{y}) = \frac{1}{P_n} \tilde{\mathcal{L}}_{\text{RN}}(\mathbf{s}, \mathbf{y})
\end{equation*}
to mitigate the dominance of samples with many positive labels, which generate more valid pairs.
For the CE loss, we normalize the labels to form a probability distribution:
\begin{equation*}
\tilde{\mathbf{y}}^i = \frac{\mathbf{y}^i}{P_n}, \quad \text{for } i = 1,\dots,n \text{ and }  
\mathcal{L}_{\text{CE}}(\mathbf{s}, \mathbf{y}) = \text{CE}(\mathbf{s} \mid \tilde{\mathbf{y}}).
\end{equation*}
We compute separate losses for clicks and orders, combining them with a weighting factor $\alpha\in(0,1)$:
\begin{equation*}
  \mathcal{L}_{\text{type}}(\mathbf{s}, \mathbf{y}_c, \mathbf{y}_o) = \alpha \cdot \mathcal{L}_{\text{type}}^c(\mathbf{s}, \mathbf{y}_c) + (1-\alpha) \cdot \mathcal{L}_{\text{type}}^o(\mathbf{s}, \mathbf{y}_o),
\end{equation*}
where type denotes the loss type either RN or CE.

\section{Experimental Setup}
We utilize a temporal train-test split for training and offline evaluation \cite{10.1145/3604915.3608839, 10.1145/3604915.3610236}. The training dataset comprises 43M and the test dataset 700k samples, both derived from anonymized user interaction logs collected from the OTTO search engine.
On this dataset, we performed extensive hyperparameter tuning for all models. The best performing DNN-based models (Two-Tower, Cross-Encoder, and Transformer) use a backbone network with $k=3$, $h=1024$, $\alpha=0.5$, $d_{\text{cat}}=128$, $d_{\text{text}}=512$ and the Adam optimizer with a batch size of 1000. The Transformer model additionally uses a single encoder layer and attention head. The Two-Tower and Cross-Encoder models are trained with a learning rate of 0.001, the Transformer model uses a learning rate of 0.0001. Dropout rates are 0.0 for the Two-Tower model, 0.3 for the Cross-Encoder, and 0.5 for the Transformer. OTTO's production baseline is a LGBM LambdaMART model trained with the LambdaRank objective, a learning rate of 0.1, ${\text{max\_depth}=12}$, ${\text{num\_leaves}=25}$ and 400 trees.

\subsection{Offline Experiments}
We evaluate model performance using NDCG for both clicks and orders, denoted as $NDCG_c$ and $NDCG_o$. This metric has shown strong explanatory power for user behavior in e-commerce settings \cite{Wang2023}. Average Item Value (AIV) serves as a proxy for revenue generated. All metrics are calculated at a cutoff of 15, which corresponds to the median scroll depth of customers on the OTTO e-commerce platform on search result pages.
Table \ref{table:offline_results} summarizes the results.
All DNNs outperform the LGBM baseline in $NDCG_c$ and AIV. However, they generally underperform in $NDCG_o$, though a few models achieve parity.
The performance of loss functions varies by architecture: the Softmax CE loss performs best for the Two-Tower model, while the RankNet loss yields stronger results for the Cross-Encoder and Transformer models. Contrary to findings in prior work \cite{Buyl2023, DBLP:journals/corr/abs-2005-10084}, the Transformer performed significantly worse in $NDCG_o$ on our dataset.
The Two-Tower model trained with Softmax CE loss emerged as the strongest candidate, improving $NDCG_c$ and AIV without degrading $NDCG_o$ and was selected for online validation.
The underperformance of all but one DNN architectures, especially the Transformer, in $NDCG_o$ might be attributed to the choice of $alpha=0.5$ and the imbalanced nature of our dataset, which is naturally skewed towards clicks.

\begin{table}[h]
  \caption{Relative improvements in $NDCG_c$, $NDCG_o$ and AIV of the Two-Tower (TT), Cross-Encoder (CR) and Transformer (TR) models over the production-grade LGBM baseline.}
  \begin{tabular}{| c | c | c | c | c | c | c |}
  \multirow{3}{*}{{\small \textbf{Metric}}} & \multicolumn{6}{c|}{\textbf{Model}} \\
  & \multicolumn{3}{c|}{\textbf{CE loss}} & \multicolumn{3}{c|}{\textbf{RN loss}} \\ 
  & \small TT & \small CR & \small TR & \small TT & \small CR & \small TR \\
  \hline
  \small $NDCG_c$ & \small +4.32\% & \small +4.32\% & \small +2.16\% & \small +3.70\% & \small \textbf{+4.63\%} & \small +3.09\% \\
  \small $NDCG_o$ & \small \textbf{0.00\%} & \small -0.30\% & \small -1.48\% & \small -1.48\% & \small -0.59\% & \small -1.78\% \\
  \small AIV & \small +2.28\% & \small +2.28\% & \small +2.89\% & \small +2.28\% & \small +2.37\% & \small \textbf{+3.16\%} \\
  \end{tabular}
  \label{table:offline_results}
  \end{table}

\subsection{Online Experiments}
To validate our offline findings, we ran an 8-week online A/B test on OTTO’s e-commerce platform. We compared the Two-Tower model with Softmax CE loss against our production-grade LambdaMART model.
As shown in Figure \ref{fig:online_ci}, the online results were in line with our offline evaluation. Using a t-test, the DNN achieved statistically significant uplifts of 1.86\%  $(p < 0.0001)$ in the total number of clicks and 0.56\% $(p < 0.01)$ in generated revenue compared to the production baseline, while units sold remained stable. The DNN has higher training and serving costs, which are negligible compared to its performance gains.

\begin{figure}[h]
  \centering
  \caption{Online A/B test results for the Two-Tower architecture with Softmax CE loss compared to our LGBM Baseline. The black bars indicate the 95\% t-test confidence interval.}
  \includegraphics[width=0.5\textwidth]{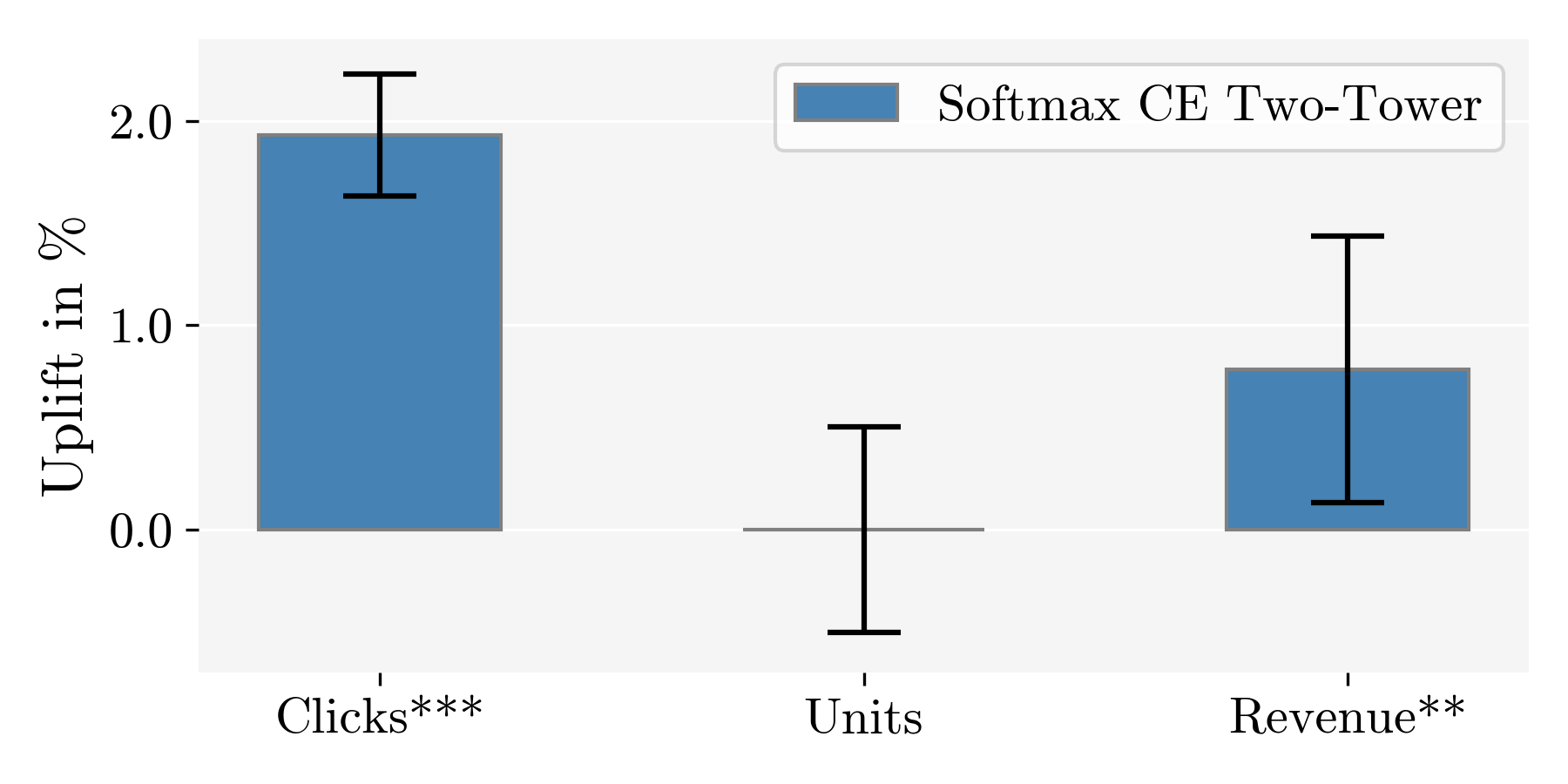}
  \label{fig:online_ci}
\end{figure}

\section{Conclusion}
This study presents a systematic comparison of DNNs and our production-grade LGBM LambdaMART model for an e-commerce LTR task. 
Using a proprietary dataset from OTTO and an 8-week online A/B test, we find that a DNN consistently outperforms a strong tree-based baseline on key engagement and monetization metrics, such as total clicks and revenue, while maintaining parity in units sold. Our results show that a simple Two-Tower model delivers competitive performance relative to more complex DNN architectures and our LambdaMART baseline. These findings suggest that deep learning approaches, when properly tuned and evaluated for production systems, can serve as a viable alternative to GBDT-based models in industrial ranking systems. Our research provides guidance to e-commerce industry practitioners who want to adopt or migrate to DNNs for their LTR tasks. 

\section{Author Bios}
\textbf{Yunus Lutz} is a Senior Data Scientist at OTTO, where he leads the development of large-scale machine learning systems powering e-commerce search.
He played a key role in developing the company’s first Learning-to-Rank model, significantly improving product search experiences. 
He is particularly interested in the connection between offline evaluation and real-world performance, helping teams build models that deliver measurable user impact.
With prior experience at Deloitte, he has a track record of delivering machine learning solutions across industries.

\textbf{Timo Wilm} is a Lead Applied Scientist at OTTO with ten years of experience, specializing in the design and integration of deep learning models for large-scale recommendation and search systems. He is responsible for translating state-of-the-art research into production-ready solutions within OTTO’s recommendation and search teams, while also contributing to industry research in the field. His work focuses on bridging the gap between academic advancements and industrial applications, ensuring that cutting-edge machine learning techniques drive measurable impact in real-world e-commerce environments.

\textbf{Philipp Duwe} is a Junior Data Scientist in the LTR Team at OTTO, dedicated to the practical use of machine learning in business settings. 
He studied Data Science and has gained quantitative experience as a working student in the finance industry.


\bibliographystyle{ACM-Reference-Format}
\bibliography{paper}

\end{document}